\documentclass[12pt, epsfig]{article}

\def\be{\begin{equation}}
\def\ee{\end{equation}}

\begin{document}


\centerline{\bf The birth of strange stars:} \centerline{\bf
kinetics, hydrodynamics and phenomenology } \centerline{\bf of
supernovae and GRBs}

\centerline{J. E. Horvath}

\centerline{\it Instituto de Astronomia, Geof\'\i sica e
Ci\^encias Atmosf\'ericas}

\centerline{\it Rua do Mat\~ao 1226, 05508-900 S\~ao Paulo SP,
Brazil}

\centerline{foton@astro.iag.usp.br}

\bigskip
\bigskip

\centerline{Abstract} We present a short review of strange quark
matter in supernovae and related explosions, with particular
attention to the issue of the propagation of the combustion in the
dense stellar environment. We discuss the instabilities affecting
the flame and present some new results of application to the
turbulent regime. The transition to the distributed regime and
further deflagration-to-detonation mechanism are addressed.
Finally we show that magnetic fields may be important for this
problem, because they modify the flame through the dispersion
relations which characterize the instabilities. A tentative
classification of explosive phenomena according to the value of
the average local magnetic field affecting the burning and the
type of stellar system in which this conversion is taking place is
presented. As a general result, we conclude that ``short''
conversion timescales are always favored, since the burning falls
in either the turbulent Rayleigh-Taylor (or even the distributed)
regime, or perhaps in the detonation one. In both cases the
velocity is several orders of magnitude larger than $v_{lam}$, and
therefore the latter is irrelevant in practice for this problem.
Interesting perspectives for the study of this problem are still
open and important issues need to be addressed.

\section{Introduction}

Intensive work in the 60´s and 70´s definitely established the
concept of elementary constituents of nucleons (quarks and
gluons). At increasing center-of-mass energy in experimental
searches of the elementary components ({\it partons}) of protons
and other hadrons revealed new physics in need of a theoretical
framework to be developed. The theory of ``new'' strong
interactions (as opposed to the ``old'' nuclear physics) was
constructed in parallel, first focused on classification schemes
(or, as is called today, flavor physics) and later on finding a
theory to describe the dynamics. The development and success of
gauge theories in the '70s eventually leaded to a non-abelian
version based on the $SU(3)_c$ symmetry group \cite{QCD} as a
``natural'' candidate for a theory of strong interactions. The
fundamental quantum number carried by the elementary constituents
(quarks) was named "color", and consequently the dynamics
involving quarks and gauge fields (gluons) become known as {\it
Quantum Chromodynamics} (QCD for short).

It was considered by many somewhat puzzling that repeated efforts
to find these entities as free particles (asymptotic states)
failed. Subsequent work elaborated on a striking feature of the
theory: that the interactions themselves preclude the appearance
of the quarks and gluons outside ordinary hadrons, they remain
confined inside them at low energies. Another property was soon
demonstrated to hold when momentum transfer scales $Q$ became
large enough. This is the so-called {\it asymptotic freedom}, and
states that the colored particles behave as if they were free in
the limit $Q \, \rightarrow \, 0$. Actually, there is an energy
(or momentum) scale above which color quantum number is not
confined any more, but how large the momentum transfer should be
(or in other words, which is energy, as measured by the
temperature or density of the ensemble allowing the deconfinement)
is still a matter of debate. These developments mean that the
early universe passed through a deconfinement $\rightarrow$
confinement phase transition along its cooling, although less
certainty holds for the  densities of the "natural" laboratories
(neutron stars) in which {\it compression} would deconfine
hadronic matter. The earliest calculations \cite{Baymetal} using
reasonable models for both the confined and deconfined phases
imprinted on successive researchers the uncertain conclusion that
quarks and gluons (forming a state known as the quark-gluon
plasma, or QGP) should appear at densities above a large
threshold, say, $10 \times \rho_{0}$; with $\rho_{0}$ the nuclear
saturation density.

From the starting of these calculations it has proved very
difficult to reliably determine the transition points, and also
the nature of the transition itself (at least when full numerical
calculations were out of sight \cite{num}). Most of the times the
conclusions had to be extracted from simultaneous extrapolations
of both a quark model, expected to be valid for $\rho \rightarrow
\infty$, and an hadronic model valid around $\rho_{0}$ but
uncertain much above it. Since there is no certainty in either
one, the final result is always subject to reasonable doubts. The
``induction'' of a definite {\it order} of the transition because
of the adopted functional forms of the thermodynamical quantities
of both sides. Nevertheless these serious and honest attempts have
proliferated until today, given that the transition is still
elusive (the extensively studied finite temperature case still has
some small uncertainty in the value of $T_{c}$ and a quite
consensual assessment of the order, see \cite{T} for details).
Recent analysis \cite{RHIC} of hadronic flows have added a lot of
excitement to these topics, since it appears that the QGP was
indeed produced in heavy ion collisions, but the asymptotic form
is {\it not} reached, rather behaving as a glass-like system.
Needless to say, this kind of studies attract a lot of attention
and offers a concrete form to glimpse the deconfined state of
hadronic matter, yet to be characterized and understood.

\section{Stable strange quark matter? }

While the study of the quark-gluon plasma occupied many studies in
connection with the early Universe and compact stars, a much
radical proposal emerged in the 80´s about it, which may be
described as follows: it is true that the asymptotic freedom
property guarantees that quarks and gluons will be the ground
state of QCD at high densities/temperatures, but it says nothing
about the ground state at lower densities or temperatures. The
everyday experience strongly suggests that ordinary hadrons
confine the quarks/gluons and thus constitute the "true" (in the
sense of $\rho \rightarrow 0$ and $T \rightarrow 0$) ground state
of hadronic matter. The emerging {\it strange matter hypothesis}
came precisely to challenge this "common sense" statement: it says
that the true ground  state of hadronic matter is a particular
form of the QGP , differing from the ordinary matter by the
presence of a key quantum number (strangeness). This is
counterintuitive to many people, but a careful look at the
physical arguments shows no inconsistency whatsoever, at least in
principle.

An argument for the SQM being the true ground state can be made as
follows: as is well-known the quantity that determines which phase
is preferred is the Gibbs free energy per particle $G/n$ as a
function of the pressure (we impose $T=0$ hereafter as appropriate
for highly degenerate hadronic matter, it is easy to see that the
term $-TS$ in the free energy disfavors SQM at high temperatures).
As $P$ is increased starting from the neighborhood of the nuclear
matter saturation point $\rho_{0}$ the asymptotic freedom
guarantees that there has to be a switch from nuclear matter (N)
to elementary hadronic constituents, that is, the lighter quarks
$u$ and $d$. The point at which this is supposed to happen will be
labelled as $P_{c}$. Thus, the doubts stated above about the
appearance of the QGP inside neutron stars may be now restated as
whether the pressure at the center is larger or smaller than
$P_{c}$.

However, this is where the concept of strangeness plays an
important role. Strangeness is the flavor quantum number carried
by $\Lambda$s and other heavy hadrons. At the elementary level, it
is carried by a different quark $s$, with current mass in the
ballpark of $\sim 100 \, MeV$, that is, light enough to be present
at a few times the nuclear saturation density. While creating
strangeness in hadrons costs energy (because strange hadrons are
heavier than non-strange ones; for instance, the $\Lambda$s are
heavier than the neutrons and so on); this trend is reversed
inside the QGP. The reason is simply the Pauli exclusion
principle: a new Fermi sea in the liquid (the one of the $s$
quark) allows a rearrangement of the energy, and sharing it lowers
the energy per particle. However, the gain is not precisely known,
but it is not impossible to imagine lowering the free energy per
particle to a value that would be lower than the mass of the
neutron $m_{n}$ even when $P \rightarrow 0$. If realized, this
would preclude the (strange) QGP to decay into ordinary hadrons
because this would {\it cost} energy and the SQM would have been
created. Put it simply, the compression would liberate the
elementary components that quickly create their own way of
surviving.  We stress that all these are bulk (i.e. large number)
concepts, and it is central to the SQM hypothesis to reach a
strangeness per baryon of the order one (and exactly one if the
strange quark had no mass to deplete its relative abundance). This
is not possible in a few-body system like a nucleus, because each
weak decay creating a strangeness unit contributes roughly with a
factor $G_{Fermi}^{2}$ to the amplitude, and thus the simultaneous
decays are strongly suppressed; this is why it has been very
difficult to produce even doubly strange nuclei, let alone higher
multiplicity ones. However, once quarks roam free in the QGP they
can easily decay by $u + d \, \rightarrow \, u + s$ because there
is plenty of phase space for the products until equilibrium is
reached. These bulk estimates have been always one way or another
behind the idea of SQM.

As it stands, the SQM hypothesis is very bold. It conjectures that
every hadron we see around us is in a metastable state, and if
conditions for creation of a large net strangeness were met, the
matter would not make back ordinary hadrons (technically it is
said that SQM constitutes a non-topological soliton stabilized
against decays by a conserved charge, the baryon number, see
\cite{FOGA} for a thorough discussion of this case and related
ones). The general idea of reaching extreme conditions and
stabilizing the QGP is already apparent in the paper of Bodmer
\cite{Bodmer}, later reintroduced and refined in references
\cite{CK, Tera, Bjo} and colorfully discussed in the paper of
Witten \cite{Ed}, which was fundamental to give a big boost to SQM
research.

SQM as a theoretical construction is interesting, but finding it
in nature would be infinitely more. Key questions of SQM such as
whether it does exist or not, and whether it has been ever
produced in the Universe are still unsolved. On the other hand, we
begin for the first time to have the possibility of falsifying
these basic questions mainly thanks to the new generation of space
telescopes (HST, Chandra, XMM) and neutrino observatories (SNO,
Kamioka, Icecube), to name just a few. These instruments may be
used to look for exotic states in compact stars and their birth
events.

Many applications of SQM in astrophysics were foreseen during the
first decade after its official birth \cite{Ed} and early infancy
\cite{FJ} Since astrophysical insight has shown to be essential in
the determination of fundamental questions related to SQM, we
shall focus briefly in a very definite (and important)
astrophysical problem, trying to give an assessment and pointing
on the uncertainties and possible directions that may be explored
in the near future. We thus restrict our discussion to SQM in
compact stars, and more specifically, on how a seed of SQM may
grow and propagate throughout a just-born neutron star. This has
been a popular choice for an energy source in GRBs and
core-collapse supernovae, therefore it is important to establish
its basic features with confidence to build on them.

\section{SQM in protoneutron stars: effects in core-collapse supernovae}

As a ``natural'' environment in which SQM might form, core
collapse supernovae has received reasonable attention \cite{BHV,
Matt, chinos, Anand}. Despite of more than three decades of
theoretical research and hard numerical modelling, the processes
that cause the explosion of massive stars are still not understood
(\cite{Bur}). If, as the more recent and detailed numerical
simulations suggest, the neutrino-driven mechanism works on
special conditions only, the current paradigm for explaining
massive star explosions would have to be deeply revised.
``Conventional'' physics has now turned attention to the role of
rotation inside the progenitor and magnetic fields \cite{SN1,
SN2}, possibly relating this problem to the GRB one
\cite{Massimo}. Although it is still too early for making
definitive conclusions, investigations including the possible
transition to deconfined QCD phases may be relevant to this
problem. The first studies of SQM in supernovae (\cite{Mac, BHV,
LBV, DPL, Anand}) showed that this hypothetic subnuclear energy
source is more than adequate to contribute to the explosion, and
that some observed characteristics in the neutrino emission of
SN1987A may be naturally explained within this scenario
(\cite{Hat, SatSuz, Mac}) (a second peak in the neutrino emission
is naturally predicted in these models, and such signal has been
tentatively associated to the late neutrinos from SN1987A detected
by Kamiokande, which have to be otherwise interpreted as a
statistical gap within the current paradigm).

From a wide perspective, supernovae are perhaps the only
astrophysical events in which we could have the possibility of
making a ``multiwavelength'' detection (neutrinos, various
electromagnetic wavelengths, gravitational waves) of the process
of SQM formation. However, these calculations are still in the
infancy, and just bold expectations have been formulated. Some
specific simulations \cite{FryerWoos} have addressed (negatively)
a few questions posed in GRB models. In addition, a firmer
detailed observational background would be needed, which
imperatively needs the occurrence of a number of supernova
explosions in the neighborhood of our galaxy, and thus is out of
any human control (in turn the instrumentation must be improved
greatly). Second, although the general picture of SQM formation in
supernovae has been qualitatively constructed, no systematic
calculations have been made. There are also many unresolved
questions related to strong interactions at high densities, which
introduce an uncomfortable degree of uncertainty in all
conclusions. We shall attempt below to describe the basics of the
SQM propagation problem, a subject that has been addressed in the
literature over the years from the kinetic/energetics point of
view \cite{Olinto, MadsenOlesen,Heiselberg,Drago,hindues}, but has
a high degree of complexity from the coupling to hydrodynamics,
much in the same way thermonuclear supernovae do. We will be
guided by the work done in the latter problem, even though most of
our discussion is new (i.e. unpublished) for the specific problem
of SQM propagation. We shall later attempt to sketch the effects
of the magnetic fields for the propagation, which leads to a
tentative classification of the different phenomenological events.

\section{SQM combustion dynamics: early stages}

As discussed and agreed in the literature, a seed of SQM must
become active or form following the standard bounce onto the
former iron core. We shall not address this problem of the seed
here, just assuming that by some of the proposed mechanisms
\cite{AFO, Mac} the seed of SQM is present within $\sim \,
seconds$ after the bounce (if the quick appearance is bypassed, a
late conversion could ensue \cite{Bombaci} but without effects in
supernovae). The neutron-SQM interface must then propagate
outwards powered by the energy release of converted neutrons, much
in the same way as a laboratory combustion.

It seems reasonable to assume the combustion to begin as a {\it
laminar} deflagration, in which the diffusion of $s$ quarks set
the scale for the flame length $l_{th}$. This has been actually
the subject of early calculations \cite{varii, Olinto,
MadsenOlesen, Heiselberg}, in which a plane front approximation
was used to obtain the laminar velocity ${\bf u}_{lam}$ as a
function of temperature, density and other relevant quantities.
The result ${\bf u}_{lam} \leq \, 10^{4} cm \, s^{-1}$ suggested
that a just-born NS would convert to a SS in $\sim \, 100 \, s$ or
so. From the combustion theory point of view this is equivalent to
decouple completely the kinetics of the burning from the
hydrodynamics of the flow in the star. Nevertheless, the
reasonable convergence of several approaches to the calculation of
${\bf u}_{lam}$ gives some confidence that the result is reliable
within the approximations.

In a situation as such (a combustion starting around the center
and propagating outwards), it has been known for many years
\cite{Darrieus, Landau} that small perturbations are unstable for
all wavelengths at a linear level. In fact Horvath and Benvenuto
\cite{HB} calculated the perturbation growth for this specific
problem with the resulting condition

\be j^{4} < 4 \sigma g \rho_{1}^{2} \rho_{2}^{2}
{1\over{(\rho_{2}-\rho_{1})}} \ee

where $j$ is the mass flux onto the flame, $\sigma$ the surface
tension, $g$ the local gravitational acceleration and $\rho_{1},
\rho_{2}$ the densities of the ``fuel'' (neutron matter) and
``ashes'' (SQM) respectively. As it stands, this is impossible to
satisfy for {\it any} deflagration (in particular, the laminar),
because by its very definition
$j^{2}=(P_{2}-P_{1})\rho_{2}\rho_{1}/(\rho_{2} - \rho_{1})$, and
thus a deflagration which must obey $P_{2} < P_{1}$ and $\rho_{2}
< \rho_{1}$ making the r.h.s a negative number. This way, the
flame wrinkles in a timescale $\leq$ the dynamical timescale
$\tau_{dyn} \sim \, 10^{-3} \, s$ (as appropriate in a
protoneutron star). Thus, the strong statement made by Landau and
Darrieus is confirmed at the linear level.

Numerical calculations of the Landau-Darrieus instability beyond
the linear level \cite{RoepkeHill} show the formation of {\it
cusps}, leading to quadratic and higher-order terms in the
dispersion relation and stabilizing the flame \cite{Zeld}. The
flame acquires a cellular shape and accelerates, since the contact
area between the fuel and ashes increases. The stationary,
scale-invariant amplitude of this cusps leads to an acceleration
of the flame, with velocity described in this regime as

\be {\bf u}_{cell} = \, {\bf u}_{lam} {\biggl(1 + 0.4(1 -
\mu)^{2}\biggr)} \ee

with $\mu = \rho_{2}/\rho_{1}$. The flame velocity is higher than
in the laminar regime by a modest amount for all reasonable
compression ratios $\mu$. An alternative cellular flame model has
been developed by Blinnikov and Sasorov \cite{BlinSas}. They
observe that the wrinkled flame can be described with a fractal
model, which in the 2-D case yields

\be {\bf u}_{cell} \simeq {\bf u}_{lam} {\biggl(
{l\over{l_{crit}}}\biggr)^{D_{cell}-2}} \ee

with $D_{cell}$ the fractal dimension of the surface and
$l_{crit}$ a suitable minimum length. A calculation of the latter
quantity finally yields

\be {\bf u}_{cell} \simeq {\bf u}_{lam} {\biggl(
{l_{max}\over{l_{crit}}}\biggr)^{0.6(1-\mu)^{2}}} \ee

where we have imposed the radial distance $l_{max}$ as the maximum
scale for which this theory is valid. Arguments related to the
propagation of L-D unstable flames suggest that $l_{crit}$ may be
identified with the {\it Markstein} length \cite{Markstein}, or at
least $\sim \, 100 l_{th}$. Eq. (4) leads to the same conclusion
as before: there is a modest acceleration of the flame and
stabilization at the small scales.

In Ref.\cite{HB}, the extreme assumption that velocity of the
flame can not become supersonic, we obtained a (small) maximum
length for this regime to hold. This should be rather interpreted
as the scale beyond which the above L-D description breaks down
definitely, due to combined additional physical effects that we
now address.

Since the gravitational pull is always being exerted onto the
flame, one could have anticipated that the cell structure can not
be scale-invariant indefinitely, and in fact disruption of the
bubbles does occur \cite{NieHill95}. A {\it turbulent cascade}
dominates the burning above certain length, which can be estimated
from the point when the velocity of turbulent fluctuations ${\bf
u'}(l)$ becomes equal to ${\bf u}_{cell}$. This defines the
so-called {\it Gibson scale} $l_{gib}$ \cite{Peters}. Imposing a
Kolmogorov spectrum (it is now established that this is more
accurate than the so-called Bolgiano-Obukhov spectrum for 3D,
whereas the latter is relevant for 2D models) ${\bf u'}(l) \propto
l^{1/3}$, it can be shown that the scaling of $l_{gib}$ is

\be l_{gib} \, \propto {\biggl( {{\bf u}_{cell} \over{{\bf
u'}(L)}}\biggr)}^{3} \ee

where the fluctuations have to be normalized to the largest scales
$L$ encountered in the system. Given that the turbulence itself
can not become supersonic (the speed of sound is already $\sim \,
c$ in the problem), and using the former value ${\bf u}_{cell}
\geq 10^{4} cm \, s^{-1}$, we obtain for $l_{gib}$ the value of
$\sim 10^{-4} cm$ as an extreme upper limit, and decreasing with
time. This is, however, initially much larger than $l_{th}$ in the
diffusive regime, and allows a classification of the burning into
the {\it flamelet} regime: while the flame propagation is still
determined by diffusion, the {\it total burning} is in turn
controlled by turbulence in a turbulent region called the {\it
flame brush}. In the flamelet regime, for all scales $\gg
l_{gib}$, the turbulent velocity ${\bf u}_{turb}$ and front width
$l_{turb}$ are determined by the Kolmogorov spectrum at the larger
scales. The important point to stress here is that the turbulent
eddy turnover controls the transport and fuel consumption, quite
unlikely a pure laminar regime \cite{Kerstein, Clavin}. Diffusion
processes do not play the dominant role once the flamelet regime
is achieved, quickly after the start of the combustion. We note
that if $l_{gib}$ decreases below the value of $l_{th}$ one can no
longer talk of a laminar regime and the burning is likely
described by the {\it distributed} regime, in which turbulent
eddies disrupt the flame and dominate the burning on macroscopic
and microscopic scales. We shall assume that the flamelet regime
exists and proceed to describe the large-scale physics, keeping in
mind the possibility of being bypassed in favor of the distributed
regime.

\section{SQM combustion dynamics: turbulent large-scale regime}

While at the small scales L-D instability affects the flame
eventually leading to the flame brush in the way described above,
on still larger scales, buoyancy of hot burned fuel (SQM)
dominates the dynamics of the process as a consequence of the
Rayleigh-Taylor instability. In fact, one obtains essentially the
R-T results by letting ${\bf u}_{lam} \rightarrow 0$ in the L-D
analysis. The classical solution of this problem \cite{Chandra} is
well-known and indicates that in the linear regime the
perturbations grow exponentially with a time scale $\tau_{RT} =
(4\pi l/g)^{1/2} (\rho/\Delta \rho)^{1/2}$, with $\Delta \rho =
\rho_{1}-\rho_{2}$. After the modes attain amplitudes similar to
the originally unperturbed, the merging/fragmentation of the
bubbles and the shear {\it Kelvin-Helmholtz} instability between
bubble surfaces give rise to a turbulent mixing layer. In models
with a single bubble scale, the velocity is

\be {\bf u}_{RT} \simeq \, {1\over{2}} {\sqrt{At \times \, l \,
\times \, g_{eff}}} \, \simeq \, {1\over{2}} {\sqrt{{1\over{2}} \,
(1-\mu) \, \times \, l \, \times \, g }} \ee

with $At$ the {\it Atwood number} and $l$ the radius of the tube
in the experiment \cite{DaviesTaylor}. In astrophysical problems a
single-scale expression is seldom enough to describe the intrinsic
multi-scale system, and a 1-D model containing most of the
relevant physics, the so-called {\it Sharp-Wheeler} model
\cite{SW} is widely used to calculate evolution of the flames. In
this picture the combustion front advances into the cold unburned
fuel with a speed of

\be {\bf u}_{SW} \, \sim \, {1\over{20}} (1 - \mu) g t \ee

It is clear that the bubble radius evolves linearly with the
distance to the center. The Sharp-Wheeler speed eq.(7) can be
identified with the effective speed of the burning provided the
latter is completed inside the R-T mixing zone.

Fractal models have also been employed as an alternative
description for the R-T regime \cite{TimWoos, Ghezzi}, with a
generic prediction that can be summarized as

\be {\bf u}_{R-T} = {\bf u}_{lam} {\biggr( {L\over{l_{min}}}
\biggl)}^{n/2} \ee

where $n=2(D-2)$ relates the index to the fractal index $D$, $L$
is the scale at which the turbulent velocities equal the R-T
instability velocity, and $l_{min}$ is the smallest scale that can
still deform the flame front (bounded from below by the Gibson
scale defined above).

For both the Sharp-Wheeler model eq.(7) and the fractal model of
eq.(8) the velocity increase is very large respect to the
``kinetic'' laminar models for the same problem. This is quite
analogous to the carbon burning regime in type I supernova models,
in which all the hydrodynamical aspects are being considered
together with the reaction kinetics. It is important to remark
that in all the cases the flames can be still defined quite
properly, and that energetics determined by Hugoniot curves are
still valid, as they should.

As performed in Type I SN studies, we plot in Fig. 1 the relevant
velocities for the burning flame as a function of the scale. From
a simple inspection of this figure, it is clear that the $n
\rightarrow SQM$ combustion should accelerate substantially when
evolving at relatively low radii (certainly $\ll \, 1 km$, still
deep in the stellar interior. Further evolution of the flame will
be discussed below, ending with some of the expected consequences
and phenomenological features.

\section{SQM combustion dynamics: distributed regime and
the transition to the detonations branch (DDT)}

The evolution of the flames described below is now quite well
established and substantiated by several numerical simulations.
One may wonder about the final outcome of the burning process when
the flame is well within the R-T stage. The possibility that
turbulence disrupts the flame on microscopic scales, which would
not be well-defined any more, can be adopted as a rough intuitive
description of the {\it distributed regime}. In the latter mixed
regions of fuel and ashes burn in regions that have a distribution
of temperatures interact strongly with the turbulence.
Alternatively, the combustion may reach the edge of the star
without reaching the distributed regime.

More rigourously, the distributed regime can be characterized by
the inequality $l_{gib} \, < \, l_{th}$. It is not clear whether
this condition is achieved in the $n \rightarrow \, SQM$
conversion. As suggested above, it may be achieved directly in the
early stages. However, and in spite that $l_{gib}$ decreases along
the propagation, {\bf u'} is clearly bounded from above by $c$.
Therefore, $l_{gib}$ may be short for the distributed regime to be
reached if it is not reached in the early stages, and this is a
point that needs a detailed investigation.

One may nevertheless entertain the possibility of a distributed
regime in the problem because it is one of the expected pathways
to the {\it detonation} branch of the combustions. In these
scenarios a detonation (self-propagated combustion mediated by a
shock) can start, for example, by means of the Zel'dovich gradient
mechanism \cite{Zel2}. For this to occur, a macroscopic region of
the mixed fuel/ashes should be able to burn ``at once'' (i.e.
allowing a supersonic phase velocity), which requires a very
shallow temperature gradient $\nabla T$. It is not known how large
the critical macroscopic region should actually be, detailed
calculations show that its value for the WD carbon burning problem
is $L_{c} \sim 10^{4} cm$, and it is likely much smaller in our
problem. Estimations of the size of the distributed flames yield
essentially $l_{dist} = \alpha l_{th} Ka$, where $\alpha$ is a
pure number and $Ka$ is the {\it Karlovitz number}, used in
turbulence studies as a measure of the quotient of diffusive to
eddy turnover time. Physically, if $l_{dist}$ can stretch to reach
the $L_{c}$ value, the system would satisfy at least a necessary
condition for a transition to detonation (since the deflagrations
come first, this is call in the literature as
Deflagration-to-Detonation Transition, or DDT, \cite{Khlo}). This
condition can be combined with the expression $l_{gib} =
l_{th}/Ka^{2}$ to yield the relation

\be \alpha^{1/3} Ka \, \geq \, {\biggl( {L_{c}\over{l_{gib}}}
\biggr)}^{1/3} \ee

converted into a bound on $\alpha$ when we observe that the
distributed regime starts at $Ka \, > \, 1$. $L_{c}$ values larger
than $\sim 10 \, cm$ would not allow the burning to become a
detonation (DDT) by the Zel'dovich gradient mechanism. Thus, a
necessary, but not sufficient, condition for the DDT can be
established whenever $L_{c} \, \leq \, 10 \, cm$.

Another condition for DDT within the gradient mechanism is related
to the hierarchy of time scales of mixing, burning and dynamical.
Contrary to the WD explosion problem, we have already seen that
$\tau_{dyn}$ is always much longer than $\tau_{burn}$ (identified
with the weak interaction time scale $\tau_{W} \, \sim 10^{-8} \,
s)$ to create the strangeness). Therefore we have

\be \tau_{mix} \leq \, \tau_{W} \, \ll \, \tau_{dyn} \ee

which yields, after substituting

\be {L_{c}\over{{\bf u'}(L_{c})}} \, \leq \, \tau_{W} \ee

using for the turbulent velocity fluctuations the estimate ${\bf
u'}(L) = (1/2) \sqrt {g_{eff}L}$ \cite{NiemWoos}, we obtain an
{\it upper} bound on $L_{c}$

\be L_{c} \, \leq \, 10^{-5} (1 - \mu) \, cm \ee

This is a small length over which to mix fluids, and would make
the former condition on the Karlovitz number (eq. 9) irrelevant,
likely leading to a DDT phenomenon immediately. Other mechanisms
for DDT do exist, but is too difficult to discuss them in
connection with our problem at this stage.

From all the above discussion we believe it is clear that the
examination of the laminar diffusive regime is just a part of the
whole very complex problem. The full evolution of the burning $n
\, \rightarrow \, SQM$ can be accurately described by using the
so-called {\it reactive Euler equations}

\be {{\partial \rho}\over{\partial t}} + {\nabla . (\rho {\bf u})}
= 0 \ee

\be {{\partial (\rho {\bf u})}\over{\partial t}} + {\nabla . (\rho
{\bf u} {\bf u})} + \nabla P = 0 \ee

\be {{\partial E}\over{\partial t}} + {\nabla . ((E+P) {\bf u})} =
0 \ee

\be {{\partial X_{i}}\over{\partial t}} + {({\bf u}.\nabla)X_{i}}
= R (T, \rho, X_{i}) \ee

with $X_{i}$ are the fraction of each quarks and the reaction
rates $R (T, \rho, X_{i})$ have to be calculated at finite
temperature for the dense environment (see, i.e. \cite{Anand}).
Due to enormously disparate length scales, ranging from $\sim$ few
$fm$ to perhaps $\sim km$, a model of the flame can facilitate the
calculations, otherwise it is known that resolving the full
structure demands a huge computational investment
\cite{WoosEtAlnew}.

\section{Role of $n \rightarrow SQM$ conversion in supernovae}

In the original proposal \cite{Mac, BHV} of a fast combustion mode
in supernovae, a newtonian calculation was employed to estimate
the dynamical quantities, in particular the energy that could be
transferred to the outer layer of a stalled shock in a massive
star. We have seen that a complex but quick sequence of phenomena
affects the flame, even if initially starts as a slow laminar
combustion. If energy can not be directly transferred to the outer
layers, SQM formation may still be important because of the
production of neutrinos by appropriate reactions in the deconfined
phase. The binding energy of the strange star has to be released
as well \cite{BomDat}, much in the same way as the binding energy
of the neutron star in the standard picture. Although new fresh
neutrinos could in principle produce a late revival of the stalled
shock wave, other features than the total released energy are
essential such as spectral features of the neutrino emission, and
more importantly (if the transition happens to be somewhat
delayed) the exact time of its occurrence, since if it occurs too
late there will be no way to explode the star by the shock
reheating mechanism at all.

While it is still not clear whether the detonation mode is
feasible, since it requires fast transport of heat to sustain the
front and a working DDT mechanism (if it is not initiated
``directly''), assuming the latter case, and since the conversion
is not expected to be exothermic all the way down to zero pressure
it is unavoidable that a detonation will become a standard shock
wave beyond some radius (assuming the MIT Bag model for SQM this
radius is the one for which $E - 3P= 4B$). This shock wave will
propagate outwards and the question is whether or not it will be
able to transfer its energy and complete the work unfinished by
the unsuccessful prompt shock wave. In turn, a more moderate
turbulent combustion (subsonic but still very fast) may be the
final outcome instead of a detonation, and its propagation would
mix the material on macroscopic scales due to the action of
Landau-Darrieus and Rayleigh-Taylor instabilities. Its role in the
reenergization of the stalled shock, possibly by neutrino
transfer, has not been calculated as yet.

A better understanding of the previous sequence of combustion
processes will also give information about the timescale of the
conversion of the star, which is closely related to the different
observational signals. These calculations also constitute an
important task for the near future.

\section{Delayed conversions, compact star structure and gamma-ray bursts}

Up to now we have considered the hydrodynamics of the reactive
flows with the assumption of its occurrence well inside the first
seconds after the prompt shock bounce. If the just-born
protoneutron stars do not collapse to black holes due to accretion
in the early stages \cite{LatPra}, and within the SQM hypothesis,
then pure strange stars, made up entirely of strange quark matter
from the center to the surface, may be the compact remnants of
supernovae. But even in the case of absolute stability, if the
transition is {\it not} triggered during the supernova explosion,
all observed ``normal'' neutron stars would be in a metastable
state, which is quite difficult to imagine because of ISM
contamination arguments \cite{AFO, Jes, MTH} and the mismatch
$\tau_{conv} \, \ll \,\tau_{star}$ between the timescale in which
favorable conditions for conversion occurs $\tau_{conv}$ and the
lifetime of the star $\tau_{star}$. According to recent
calculations the deconfinement transition is more likely to occur
by heating and compression during the Kelvin-Helmholtz phase of
proto-neutron star (PSN) evolution (see, for instance,
\cite{LuBe}). If it did not happen there, once the PNS has cooled
to temperatures below $\sim \, 1 MeV$, only accretion from a
companion star or strangelet contamination would allow the
transition (and many barriers may preclude its occurrence), even
in the case where it is energetically favored. Thus, the existence
of strange stars is determined not only by fundamental questions
concerning the true ground state of dense matter but also by the
exact physical conditions in the specific astrophysical
environments together with the plausibility of the conversion
mechanisms in these situations.

The SQM conversion has been repeteadly associated with gamma-ray
bursts. Many works in the past have explored the idea that the
conversion of NM into SQM in NSs may be an energy source for GRBs
(\cite{AFO1, MX, Hae, CD, BomDat, Chinos2, Rach, PacHaens,
Berezhiani}). These models mostly address spherically symmetric
conversions of the whole NS rendering isotropic gamma emission.
Accumulating observational evidence suggests that at least
``long'' GRBs are strongly asymmetric, jet-like outflows, a
feature that needs some crucial ingredient in the SQM physics
formation/propagation to proceed. The ``short'' burst subclass was
not obviously asymmetric prior to HETE2 and SWIFT data, but now
evidence has mounted for a substantial asymmetry (but not extreme)
in them. The association of Type Ib/c with a few GRBs has
reinforced the investigation of underlying explosion mechanisms,
and the absence of a temporal break in most of the light curves
(interpreted in terms of a collimated jet effect) is a puzzling
feature that might be related to the total energy budget in a yet
unclear resolution.

A new potentially important feature recently recognized in this
class of models is that if a conversion to SQM actually begins
near the center of an NS, the presence of a moderate magnetic
field B ($ \sim 10^{13}$ G) will originate a prompt {\it
asymmetric} gamma emission, which may be observed as a short,
beamed GRB after the recovery of a fraction of the neutrino energy
via $\nu {\bar\nu} \rightarrow e^{+}e{-} \rightarrow \gamma
\gamma$ \cite{Lug}. The basic physical effect is again related to
the instabilities described in the former sections: the influence
of the magnetic field expected to be present in NS interiors
quenches the growth of the hydrodynamic instabilities in the
equatorial direction of the star (parallel to the magnetic field)
while it allows them to grow in the polar one. As a result, the
flame will propagate much faster in the polar direction, and this
will result in a strong (transitory) asymmetry in the geometry of
the just formed core of hot SQM, which will resemble a cylinder
orientated in the direction of the magnetic poles of the NS. While
it lasts, this geometrical asymmetry gives rise to a bipolar
emission of the thermal neutrino-antineutrino pairs produced in
the process of SQM formation. This is because almost all the
thermal neutrinos generated in the process of SQM formation will
be emitted in a free streaming regime through the polar cap
surface, and not in other directions due to the opacity of the
matter surrounding the cylinder. The neutrino-antineutrino pairs
annihilate into electron-positron pairs just above the polar caps
of the NS, giving rise to a relativistic fireball, thus providing
a suitable form of energy transport and conversion to
gamma-emission that may be associated to short gamma-ray bursts. A
unifying scheme in which SQM appearance produces spherical
ejection phenomena to highly asymmetric gamma beaming, as a more
or less continuous function of the magnetic field $B$ and the
astrophysical system under examination may be possible, and is
tentatively sketched in Table 1.
\bigskip

\noindent {\bf Table 1.} Tentative classification of explosive
events due to SQM in several stellar systems

\bigskip
\noindent
\begin{tabular}{|cccc|r|}
    \hline
 Mag. field (G) &Type II SN &  LMXB-HMXB$^{\ast}$ & AIC(?)$^{\dagger}$    \\
    \hline
 $0 < B < 10^{12}$& "normal" SN&spherical,weak short GRB&UV-X flash \\
 $B \sim 10^{13}$& bipolar SN &bipolar,strong short GRB&bipolar UV-X flash\\
 $ B \geq 10^{14}$ & ? &jet-like,weak short GRB&jet-like UV-X flash\\
 $B \gg 10^{15-16}$ &  -- & -no SQM formation- & --\\

    \hline
\end{tabular}

\bigskip
\noindent $\ast$ only if $NM \rightarrow SQM$ conversion is
sometimes suppressed when a NS is formed.

\noindent $\dagger$ upper limit to the rate $\sim 10^{-4} yr^{-1}
galaxy^{-1}$ needs to be revised if SQM burning occurs modifying
nucleosynthetic yields.

\bigskip

We are still very far from a thorough understanding of magnetic
field effects, and a reliable simulation is even more challenging
than simulating the $B=0$ reactive Euler equations (13-16).
However, we believe that it is fair to say that magnetic fields
are relevant for the physics of the conversion, even at moderate
values. In summary, we may be witnessing an ultimate subnuclear
energy source in action, powering SN-GRBs if SQM exists.

\section{Conclusions}

We have presented a discussion of the main features of hypothetic
$n \rightarrow SQM$ conversions inside neutron stars. We have
shown that even if the initial state of the process could be a
laminar deflagration, the hitherto ignored hydrodynamic
instabilities (Landau-Darrieus and Rayleigh-Taylor) quickly take
over and determine the propagation through the vast majority of
the star, in a regime of turbulent deflagration \cite{HB, Lug}.
Models which ignore hydrodynamics altogether or just concentrate
on the energy conditions to determine the combustion mode miss
completely this important features. In particular, the association
of long timescales (up to $10^{3}-10^{4} \, s$) of GRBs based on
the identification of a laminar deflagration as the relevant
timescale in the process is not tenable \cite{HZu}. Other proposed
models differ in their kinetic aspects, for example, models in
which energy is obtained by pairing quarks \cite{LH2, LH3, Hsu,
Ouy} typically operate on strong interaction timescales, and thus
may be thought as an isocoric burning, i.e. much faster than the
described instability scenario. Still other energy transfer
mechanisms are possible \cite{Xu}, and certainly the issue of
neutrino transport from the reaction zone ahead has never been
addressed in detail \cite{BenLug}, although there is plenty of
energy carried by them.

It is still possible that all these regimes could be bypassed in
favor of a ``prompt'' detonation mode started at the very central
region, for example, by the sudden conversion of a macroscopic
small region, further sending a shock wave with $\sim$ half of the
initial overpressure \cite{Mamaia}. Propagation of such a
combustion mode is in principle possible \cite{Mac, Tokareva}, but
more detailed studies have yet to be performed on this problem by
coupling properly the energy transport to the structure of the
flame front. Models treating the conversion much in the same way
as a plain phase transition are even more remotely relevant to the
actual physics.

\section{Acknowledgements}

Along several years of SQM research many people contributed to
clarify and explain features of the physics and astrophysics of
SQM to the author, and provided guidance in many respects. Among
them we wish to acknowledge O.G. Benvenuto, H. Vucetich,  G.
Lugones, J.A. de Freitas Pacheco and I. Bombaci, colleagues and
friends. The authors wish to thank the S\~ao Paulo State Agency
FAPESP for financial support through grants and fellowships, and
the partial support of the CNPq (Brazil).

\clearpage

\begin{figure}
\caption{Scales in the SQM burning problem. }

At a given instant the regimes dominating the burning are shown as
a function of the lengthscale. Below $\sim \, 100 \, l_{th}$ the
laminar flame ensues. Cells appear above that scale and produce a
weakly-dependent velocity (as described by fractal models, for
instance). Above $l_{gib}$ cellular stabilization fails and above
a transition scale the buoyancy ultimate dominates the burning
${\bf u_{RT}} \propto \, l^{1/2}$. It should be kept in mind that
the distributed regime may be directly reached, disrupting the
flame that no longer follows the regimes of Fig.1
\end{figure}

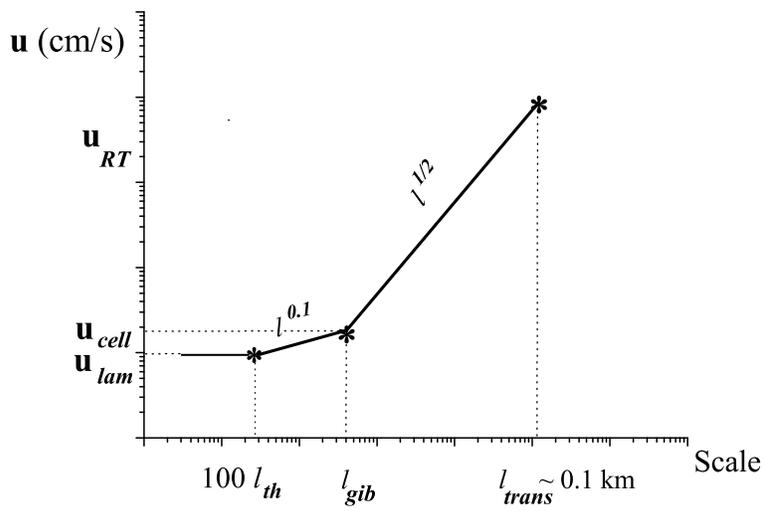
\end{document}